\documentclass{article}

\usepackage{natbib}
\usepackage{algorithmic}
\usepackage{algorithm}
\setcitestyle{numbers,square}
\usepackage[preprint]{neurips_2023}

\usepackage[utf8]{inputenc} 
\usepackage[T1]{fontenc}    
\usepackage{hyperref}       
\usepackage{url}            
\usepackage{booktabs}       
\usepackage{amsfonts}       
\usepackage{nicefrac}       
\usepackage{microtype}      
\usepackage[table]{xcolor}         
\usepackage{graphicx}
\usepackage{multicol}
\usepackage{multirow}
\usepackage{makecell}
\usepackage{listings}
\usepackage{bm}
\usepackage{subfig}

\title{Boosting Fast and High-Quality Speech Synthesis with Linear Diffusion}

\author{%
  Haogeng Liu\textsuperscript{1, 2}, Tao Wang\textsuperscript{1, 2}, Jie Cao\textsuperscript{1}, Ran He\textsuperscript{1, 2}\footnote{Ran He is the corresponding author.}, Jianhua Tao\textsuperscript{3}\\
 \textsuperscript{1}Institute of Automation, Chinese Academy of Sciences\\
 \{liuhaogeng2022,wangtao2018\}@ia.ac.cn, jie.cao@cripac.ia.ac.cn, rhe@nlpr.ia.ac.cn\\
  \textsuperscript{2}University of Chinese Academy of Sciences\\
  \textsuperscript{3}Department of Automation, Tsinghua University \\
   jhtao@tsinghua.edu.cn\\}

\begin{document}

\maketitle

\begin{abstract}
Denoising Diffusion Probabilistic Models have shown extraordinary ability on various generative tasks. However, their slow inference speed renders them impractical in speech synthesis. This paper proposes a linear diffusion model (LinDiff) based on an ordinary differential equation to simultaneously reach fast inference and high sample quality. Firstly, we employ linear interpolation between the target and noise to design a diffusion sequence for training, while previously the diffusion path that links the noise and target is a curved segment. When decreasing the number of sampling steps (i.e., the number of line segments used to fit the path), the ease of fitting straight lines compared to curves allows us to generate higher quality samples from a random noise with fewer iterations. Secondly, To reduce computational complexity and achieve effective global modeling of noisy speech, LinDiff employs a patch-based processing approach that partitions the input signal into small patches. The patch-wise token leverages Transformer architecture for effective modeling of global information. Adversarial training is used to further improve the sample quality with decreased sampling steps. We test proposed method with speech synthesis conditioned on acoustic feature (Mel-spectrograms). Experimental results verify that our model can synthesize high-quality speech even with only one diffusion step. Both subjective and objective evaluations demonstrate that our model can synthesize speech of a quality comparable to that of autoregressive models with faster synthesis speed (3 diffusion steps).

\end{abstract}

\section{Introduction}
Deep generative models have made tremendous strides in the realm of speech synthesis. Overall, contemporary speech synthesis techniques can be broadly categorized into two paradigms: methods that leverage likelihood-based modeling and methods based on generative adversarial networks (GANs). For example, WaveNet \citep{oord2016wavenet}, an autoregressive likelihood-based model, can synthesize high-quality speech. However, it is also characterized by expensive computational cost at inference time. Moreover, alternative approaches such as flow-based model \citep{prenger2019waveglow} and Variational AutoEncoders (VAE) \citep{kingma2019introduction} have their own limitations on sample quality. While GAN-based models \citep{goodfellow2020generative,kumar2019melgan,kong2020hifi} exhibit fast-paced speech synthesis, they are concurrently beset by training instability and limited sample diversity.

An emerging group of generative models, Denoising Diffusion Probabilistic Models (DDPMs) \citep{ho2020denoising,song2020score}, a likelihood-based model, have become increasingly popular in speech synthesis. For instance, WaveGrad \citep{chen2020wavegrad} and DiffWave \citep{kong2020diffwave} produce high-quality samples that match the quality of autoregressive methods. However, the iterative optimization in DDPMs significantly slows down sampling speed. To tackle this problem, existing approaches either design an extra structure such as a noise schedule network \citep{lam2022bddm, huang2022fastdiff} or describe the random diffusion process with the ordinary differential equation (ODE)~\citep{liu2022flow}. However, ODE-based diffusion still needs several steps to produce high-fidelity sample.

Inspired by rectified flow \citep{liu2022flow}, We proposed a conditional diffusion model. During the training process, we performs a linear interpolation between a target sample and initial standard Gaussian noise to construct the diffusion sequence and train the denoising network to fit the path reversely. 
For the inference process, we reconstruct the target through the Euler sampling method from a randomly sampled standard Gaussian noise.

In the light of the success of Vision Transformer (ViT) \citep{bao2022all} for image synthesis, we propose a similar structure for audio that turns continual sampling points into an audio patch and apply Transformer \citep{vaswani2017attention} to build contextual connections for these tokens. We then use a Time-Aware Location-Variable Convolution \citep{huang2022fastdiff} module for fine-grained detail restoration. As demonstrated in previous work \citep{xiao2022tackling}, the combination of DDPMs and GANs has shown promising performance. In line with this, we incorporate adversarial training into our method to enhance the quality of generated samples while reducing the number of required iteration steps.

Overall, the main contributions of this paper are:
\begin{itemize}
\item A linear conditional diffusion algorithm with an ordinary differential equation is proposed to reduce the steps required during inference. Experiments demonstrate this method can synthesis relatively high-fidelity speech with limited steps.
\item A Transformer-based architecture for audio denoising is introduced. As Transformer captures long in-context information efficiently, it enables rapid enlargement of the receptive field. To the best of our knowledge, we are the first to apply Transformer for conditional waveform generation (i.e., vocoder).
\item Adopting implicit diffusion to combine Linear diffusion with adversarial training for further reducing the steps for inference while maintaining the generated speech's high quality. Experiments show the introduction of adversarial training enables the proposed model to synthesize relatively high-fidelity speech even with only 1 step.
\end{itemize}

\section{Background}
Denoising Diffusion Probabilistic Models (DDPMs) belongs to the likelihood-based generative models that have advanced the development of speech synthesis and image generation \citep{rombach2022high}. The main idea of DDPMs is to build a diffusion sequence and train a denoising network for reversing the diffusion process iteratively. Given the origin data sample $\mathbf{x}_0$, the forward process is formalized as a Markov chain using a stochastic differential equation to model:
\begin{equation}
\mathrm{d}\mathbf{x}_t = \mathbf{\boldsymbol{\mu}}(\mathbf{x}_t, t)\mathrm{d}t + \sigma(t)\mathrm{d}\mathbf{w}_t,
\end{equation}
where $t\in [0, T], T > 0$ is a manually set parameter, $\mu(\cdot, \cdot).$ and $\sigma(\cdot)$ are the drift and diffusion coefficients. Besides $\mathbf{w}_t$ denotes the standard Brownian motion~ \citep{song2023consistency}. 

The forward process can be defined as:
\begin{equation}
q_(\mathbf{x}_{t}|\mathbf{x}_{t-1}) = N(\mathbf{x}_t;\sqrt{1-\beta_t}\mathbf{x}_{t-1}, \beta_t\boldsymbol{I}).
\end{equation}

the reverse denoising process is defined as:
\begin{equation}
p_\theta(\mathbf{x}_{t-1}|\mathbf{x}_t) = N(\mathbf{x}_{t-1};\boldsymbol{\mu}_\theta(\mathbf{x}_t, t),\sigma^2\boldsymbol{I}).
\end{equation}
The training object of the model is to match the true denoising distribution $q(\mathbf{x}_{t-1}|\mathbf{x}_{t})$ with the parameterized denoising model $p_\theta(\mathbf{x}_{t-1}|\mathbf{x}_{t})$. 

It has been shown that DDPMs are capable of learning diverse data distributions in various domains. However, the iterative optimization process for high-fidelity sample generation requires up to thousands of steps. With such a large number of reverse steps, the assumption that $p_\theta(\mathbf{x}_{t-1}|\mathbf{x}_t)$ follows a Gaussian distribution can be met. When reducing the number of reverse steps, $p_\theta(\mathbf{x}_{t-1}|\mathbf{x}_t)$ would become a more complex distribution. To reduce the required number of steps, some works \citep{xiao2022tackling} proposed combining DDPMs with GAN, utilizing GAN's complex distribution modeling ability to train the reverse process with fewer steps. Their experiments showed that DDPMs improves the training stability of GAN while GAN accelerates the reverse diffusion process, allowing the model to generate high-quality samples with only a few steps. This method was then applied to text-to-Mel-spectrogram generation \citep{liu2022diffgantts}. On the other hand, some works\citep{liu2022flow} built the diffusion path using linear interpolation, dropping the strict assumptions of DDPMs, and were able to produce high-fidelity samples with limited steps (even with one single diffusion step).

\section{Method}
In this section, we present a comprehensive overview of our proposed methodology. We commence by introducing the fundamental concept of linear diffusion and subsequently elaborate on the formulation of our algorithm. Lastly, we provide a detailed exposition of the structure of LinDiff.

\subsection{Linear Diffusion}
Different from the origin DDPMs, this work uses an ordinary differential equation (ODE) to model the diffusion process. The proposed diffusion process can be described by the following formula:
\begin{equation}
\mathrm{d}\mathbf{a}_t = \mathbf{v}_d\cdot{\mathrm{d}t}, t\in[0, T),
\end{equation}
where $\mathbf{a}_t$ denotes the noisy audio at time t and $\mathbf{v}_d$ can be computed as follows:
\begin{equation}
\mathbf{v}_d = \mathbf{a}_T - \mathbf{a}_0,
\end{equation}
where $\mathbf{a}_0$ denotes the origin audio and $\mathbf{a}_T$ denotes the noise sampled from standard Gaussian distribution. Alternatively, we employ $\mathbf{a}_t^{rev}$ to represent the noise audio in reverse process. This could directly be computed given time step $t$ in reverse process:
\begin{equation}
\label{eq4}
\mathbf{a}_t^{rev} = (1 - \frac{t}{T})\mathbf{a}_T + \frac{t}{T}\mathbf{a}_0.
\end{equation}

For the reverse process, the step $t$ is the opposite of that in diffusion process. The formulation are is listd below:
\begin{equation}
\mathbf{a}_T^{rev} = \int_{0}^{T}{v}(\mathbf{a}_t^{rev}, c, t)\mathrm{d}t.
\end{equation}

The Euler method is applied to solve this equation in an iterative manner as follows:
\begin{equation}
\mathbf{a}_{t+1}^{rev} = \mathbf{a}_{t}^{rev} + {v}(\mathbf{a}_t^{rev}, \mathbf{c}, t)\frac{1}{T}, t\in[0, T),
\end{equation}

The implicit reverse diffusion process in the paper \citep{song2020denoising} is adopted here. So LinDiff predicts the blurry target and then computing the direction ($v$), assuming LinDiff as function $f(\cdot)$: 
\begin{equation}
\label{eq9}
{v}(\mathbf{a}_t^{rev}, \mathbf{c}, t) = {f}(\mathbf{a}_t^{rev}, \mathbf{c}, t) - \mathbf{a}_T.
\end{equation}

The diffusion loss here is:
\begin{equation}
\label{diffloss}
\mathcal{L}_{diff} = \mathbb{E}[({f}(\mathbf{a}_t^{rev}, \mathbf{c}, t) - \mathbf{a}_0)^2].
\end{equation}
Though it seems that the proposed model could generate samples at any time $t$ with one step, our experiment demonstrates that generating samples with more steps in an iterative manner could produce higher-quality speeches though we sacrificed efficiency. It is the main difference that LinDiff differs from GAN. Besides, by directly predicting the origin sample, we could incorporate adversarial training in our training, which improves the model's performance.

\subsection{LinDiff}
LinDiff is a backbone that combines Transformer \citep{vaswani2017attention} and Convolutional Neural Network for diffusion-based speech synthesis. We also apply discriminators for the training process. Specifically, LinDiff parameterizes the noise prediction network in Eq. \ref{eq9}. It takes the diffusion step $t$, the condition $\mathbf{c}$ and the noisy audio $\mathbf{a}_t^{rev}$ as inputs and predict the target speech $\mathbf{a}_{gt}$. 

\paragraph{Audio Transformer block} Inspired by the U-ViT backbone in diffusion models \citep{bao2022all}, we introduce an Audio Transformer (AiT) block for speech synthesis. To achieve this, we partition the input noise into smaller patches, treating each patch as a token. Subsequently, we apply a linear transformation to obtain patch embeddings, which are then fed into the Audio Transformer block. In our experimentation, we explore various patch sizes and find that reducing the patch size improves model performance. However, it is important to note that this improvement comes at the expense of increased computational cost due to the higher number of patches. Thus, a trade-off between model performance and computational efficiency must be carefully considered when selecting the optimal patch size. In our implementation, we set the patch size to be 64.

\paragraph{Feature fusion} For each time step $t$, we follow paper \citep{zeng2021lvcnet} to embed the step into an 128-dimensional positional encoding vector $\mathbf{e}_t$ and then apply linear transformation to turn it into diffusion-step embedding $\mathbf{t}_{emb}$. We propose Time-Adaptive Layer Norm to fuse the step information. Supposing the noise audio feature is $\mathbf{x}$ and  $\mathrm{LN}$ denote layer normalization \citep{ba2016layer}.
\begin{equation}
\mathrm{TALN}(\mathbf{x}, \mathbf{t}_{emb}) = g(\mathbf{t}_{emb})\cdot\mathrm{LN}(\mathbf{x}) + b(\mathbf{t}_{emb}).
\end{equation}
To fuse the accoustic feature, a cross attention module was employed. We first use linear transformation to turn it into hidden features. And then the input noisy audio feature sequence will assume the role of query, interacting with the Mel hidden feature sequence. 

\paragraph{Post Conv}
Due to partitioning the input sequence into small patches, the resolution of the model is limited and the output sequence obtained through AiT losses much high-frequency information. However, as the human ear is sensitive to speech, we here add a Post Convolution module to process the details of the output. We follow \cite{huang2022fastdiff} to use a Time-Aware Location-Variable Convolution module with some simple Conv1d layer as our Post Conv module. Experimental results demonstrate that this approach improves the quality of the audio.

\begin{figure}
  \centering
  \includegraphics[width=14cm]{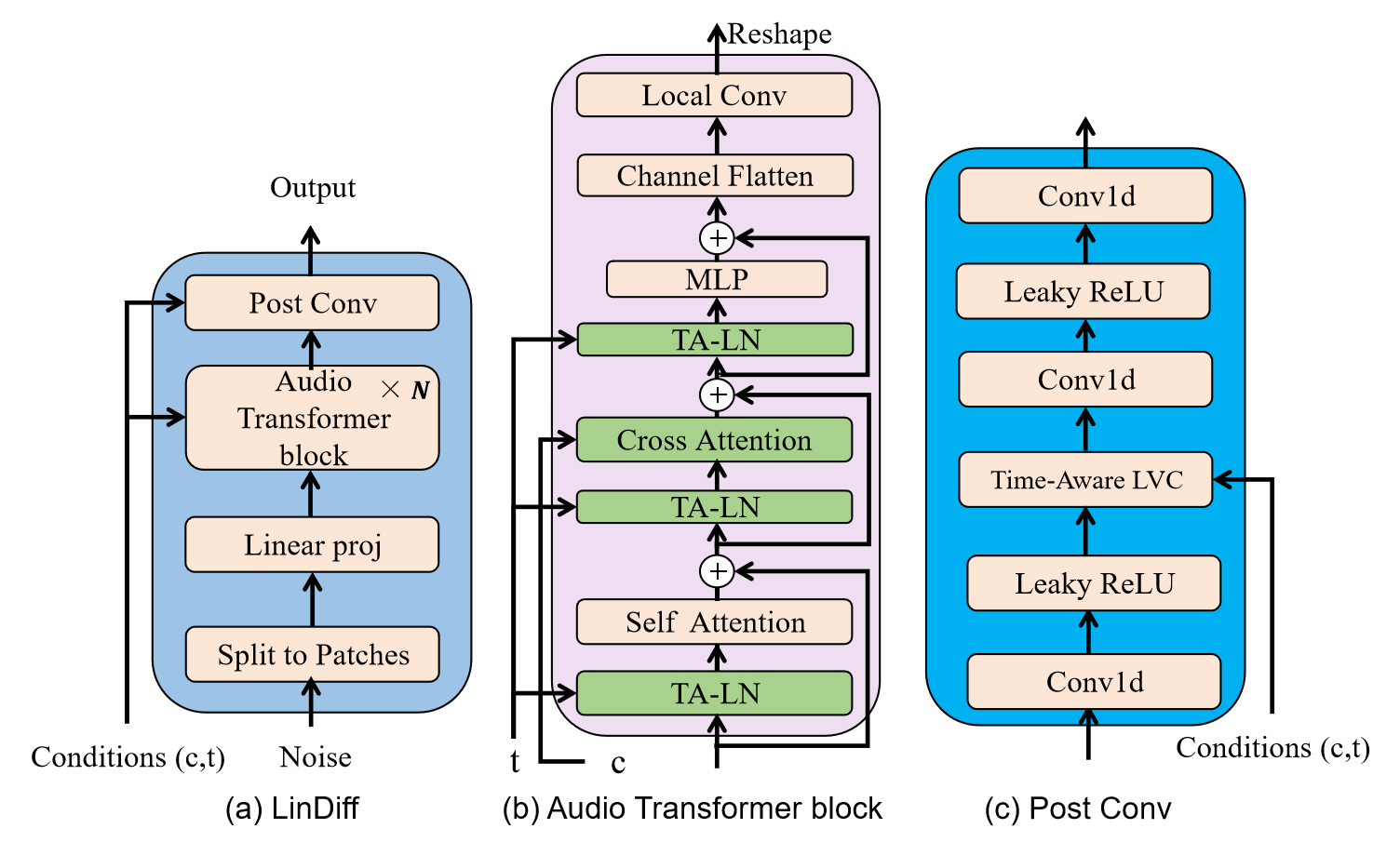}
  \caption{The overall architecture of LinDiff. TA-LN represents TimeAdaptive LayerNorm. Time-Aware LVC represents Time-Aware Location-Variable Convolution module proposed by \cite{huang2022fastdiff}.}
  \label{fig:figure1}
\end{figure}
\subsection{Training loss}
The aim of this work is to propose a model with fast inference speed, high-quality speech generation capability and training stability. However, reducing the number of sampling steps in our models led to a decrease in the quality of generated speech, similar to the findings in existing literature. To address this issue, we draw inspiration from the DiffGAN~\citep{xiao2022tackling} and introduce the adversarial training scheme into our model. This allows us to maintain high-quality speech while reducing the number of iterations. Additionally, the presence of the diffusion process improves the stability of adversarial training.
For the existing work, \citep{xiao2022tackling} parameterize the denoising function as an implicit denoising model. We follow this way. Specifically, instead of calculating $\mathbf{a}_{t+1}^{rev}$ directly from $\mathbf{a}_t^{rev}$, we first predict $\mathbf{a}_T^{rev}$\footnote{For the reverse process, we set the starting step at Gaussian noise to be 0.} (The target waveform) and then obtain $\mathbf{a}_{t+1}^{rev}$ with following formulation:
\begin{equation}
\label{eq12}
\mathbf{a}_{t+1}^{rev} = \mathbf{a}_{t}^{rev} + (\mathbf{a}_{T}^{rev} - \mathbf{a}_{0}^{rev})\times \frac{1}{T}.
\end{equation}
Our discriminator is denoted as $D_{\phi}(\mathbf{a}_{T}^{rev})$, where $\mathbf{a}_T^{rev}$ denotes the predicted blurry audio. 
We apply three discriminators from different perspectives on $\mathbf{a}_{T}^{rev}$ predicted from each time step. One of the discriminators examines the samples from a spectral perspective, while the other two adopt the multi-scale and multi-period discriminators utilized in HiFiGAN~\citep{kong2020hifi}, both of them perform discrimination in the time domain.

Our loss consists of three parts: diffusion loss, frequency-domain reconstruction loss, and adversarial loss. 

Diffusion loss:
\begin{equation}
\mathcal{L}_{diff} = \mathrm{MSE}(\mathbf{wav}_{gt} - \mathbf{a}_{T}^{rev}).
\end{equation}
We use $\mathrm{STFT}$ to represent Short-time Fourier Transform. Then Frequency-domain reconstruction loss:
\begin{equation}
\mathcal{L}_s = \mathrm{MSE}(\mathrm{STFT}(\mathbf{wav}_{gt}) - \mathrm{STFT}(\mathbf{a}_{T}^{rev})),
\end{equation}
We set four pairs of ($window\_size,hop\_length,n\_fft$) and randomly choose one of them to carry out the STFT to alleviate overfitting issue.

We assume $D_{\phi}(\mathbf{a}_{T}^{rev})$ to represent the discriminator. For the generator, adversarial loss:
\begin{equation}
\mathcal{L}_{adv}^g = (1 - D_{\phi}(\mathbf{a}_{T}^{rev}))^2.
\end{equation}
We use $\mathrm{sg(\cdot)}$ to represent stop gradient, then for the discriminator:
\begin{equation}
\mathcal{L}_{adv}^d = D_{\phi}^2(\mathrm{sg}(\mathbf{a}_{T}^{rev})) + (1 - D_{\phi}(\mathbf{a}_{T}^{true}))^2.
\end{equation}
Total loss for generator is:
\begin{equation}
\mathcal{L}_{gen} = \mathcal{L}_{adv}^g + \mathcal{L}_s + \mathcal{L}_{diff}.
\end{equation}
It is worth mentioning that our discriminator is composed of three sub-discriminators, resulting in a relatively high computational cost that slows down the training process. In order to expedite the training, the discriminator is updated every 5 training steps. To maintain a balanced adversarial training process for the generator, a weight of 0.2 was added on its adversarial loss.
\subsection{Algorithm}
We present a concise summary of the pseudocode outlining the training and inference processes of our model. The training process \ref{alg1} consists of three stages. In the first stage, we sample short audio clips and train the model without adversarial training. In the second stage, we introduce adversarial training while still using short audio clips, and we update the discriminator's weights at each step. In the third stage, we sample long audio clips for training. Regarding the inference process \ref{alg2}, we begin by randomly sampling noise from a Gaussian standard distribution. Assuming the Mel spectrogram's size is $T \times 80$, the sampled noise has a size of $(256 \times T) \times 1$, which corresponds to the result of the Short-Time Fourier Transform (STFT) applied during the generation of the training dataset.
\begin{algorithm}
\label{alg1}
\renewcommand{\algorithmicrequire}{\textbf{Input:}}
\renewcommand{\algorithmicensure}{\textbf{Output:}}
\caption{LinDiff Training Algorithm}
\label{alg1}
\begin{algorithmic}[1]
\STATE \textbf{Input}: Speech and correlated Mel-spectrograms ($\mathbf{a}_{T}, \mathbf{condition}$), Random Gaussian noise $\mathbf{a}_0$, Total diffusion steps $T$
\REPEAT
\STATE Randomly select a step $t\in[0,T)$
\STATE Calculate $\mathbf{a}_t$ with sampled $t$, noise $\mathbf{a}_0$ and ground truth wavform $\mathbf{a}_T$ according to Eq \ref{eq4}
\STATE Predict the target $\mathbf{a}_T^{rev} = f(\mathbf{a}_t, \mathbf{condition}, t)$
\IF{Stage 2}
\STATE Calculate the discriminator's loss $\mathcal{L}_{adv}^d=D^2_{\phi}(\mathrm{sg}(\mathbf{a}_T^{rev}))+(1-D_\phi(\mathbf{a}_{T}))^2$
\STATE Perform backpropagation on $\mathcal{L}_{adv}^d$ and update the weights of the discriminator
\STATE Calculate LinDiff's loss $\mathcal{L}_{gen}=(1-D_\phi(\mathbf{a}_T^{rev}))^2 + \mathrm{MSE}(\mathbf{a}_T^{rev}, \mathbf{a}_T)+\mathrm{MSE}(\mathrm{STFT}(\mathbf{a}_T^{rev}), \mathrm{STFT}(\mathbf{a}_T))$
\STATE Perform backpropagation on $\mathcal{L}_{gen}$ and update the weights of LinDiff
\ENDIF
\IF{Stage 3}
\IF{$Step \% 5 == 0$}
\STATE Calculate the discriminator's loss $\mathcal{L}_{adv}^d=D^2_{\phi}(\mathrm{sg}(\mathbf{a}_T^{rev}))+(1-D_\phi(\mathbf{a}_{T}))^2$
\STATE Perform backpropagation on $\mathcal{L}_{adv}^d$ and update the weights of the discriminator
\ENDIF
\STATE Calculate LinDiff's loss $\mathcal{L}_{gen}=(1-D_\phi(\mathbf{a}_T^{rev}))^2 + \mathrm{MSE}(\mathbf{a}_T^{rev}, \mathbf{a}_T)+0.2\times \mathrm{MSE}(\mathrm{STFT}(\mathbf{a}_T^{rev}), \mathrm{STFT}(\mathbf{a}_T))$
\STATE Perform backpropagation on $\mathcal{L}_{gen}$ and update the weights of LinDiff
\ENDIF
\UNTIL convergence
\end{algorithmic}
\end{algorithm}

\begin{algorithm}[!h]
\label{alg2}
	\renewcommand{\algorithmicrequire}{\textbf{Input:}}
	\renewcommand{\algorithmicensure}{\textbf{Output:}}
	\caption{LinDiff inference algorithm}
	\label{alg2}
	\begin{algorithmic}[1]
	    \STATE \textbf{Input}: Random noise $\mathbf{a}_0$, Mel-spectrogram ($\mathbf{condition}$), Total diffusion steps $T$
        \STATE Set $t = 0$
	    \FOR{$t < T$}
        \STATE Predict blurry target $a_T^{rev} = f(\mathbf{a}_t, \mathbf{condition}, t)$
		\STATE Caculate $\mathbf{a}_{t+1}^{rev}$ with $\mathbf{a}_{t}^{rev}, \mathbf{a}_T^{rev}$ and $\mathbf{a}_0$ according to Eq \ref{eq12}
        \STATE $t = t + 1$
        \ENDFOR
        \RETURN $\mathbf{a}_T^{rev}$
	\end{algorithmic}  
\end{algorithm}

\section{Experiments}

\subsection{Setup}
\paragraph{Datasets}
In this study, we evaluated the proposed model on two distinct datasets.\footnote{Demos can be found at https://liuhaogeng.github.io/LinDiff/} The first dataset is the LJ Speech dataset \citep{ito2017lj}, which is composed of 13,100 audio clips at a sampling rate of 22050 Hz, spoken by a single speaker reading passages from 7 non-fiction books. This dataset spans approximately 24 hours of audio in total. The second dataset is the LibriTTS dataset \citep{zen2019libritts}, which contains 585 hours of speech data from 2484 speakers. In all of the experiments, we utilized a 16-bit, 22050 Hz sampling rate. For the speech synthesis task, we used 80-band Mel-spectrograms as the condition. These spectrograms were extracted using Hann windowing with a frame shift of 12.5-ms, frame length of 50-ms, and a 1024-point Fourier transform.

\paragraph{Model Configurations}
The LinDiff model comprises three key components: a patch-embedding module, an Audio Transformer, and a Post-Conv module. Specifically, we utilize a patch size of 64 and apply a linear transformation to the patch-embedding module, which generates a 256-dimensional embedding. The Audio Transformer component of the model consists of four layers, with each layer having a hidden dimension of 256 and four attention heads for both self-attention and cross-attention. The MLP within the Audio Transformer layers utilizes Conv1d. The Post-Conv module uses Conv1d and Time-Aware Location-Variable Convolution with 32 channels.

\paragraph{Training and Evaluation}
For this particular experiment, we trained the LinDiff model until it reached 200k steps using the Adam optimizer \citep{kingma2017adam} with $\beta_1 = 0.9, \beta_2 = 0.98, \epsilon = 10^{-9}$. Both models were trained on 4 NVIDIA GeForce RTX 3090 GPUs, using randomly sampled audio clips that matched the maximum transformer length (we set it 3600, which means max audio length is $3600*64/22050=10.44$ s), with a total batch size of 16. Initially, we trained the model for 10k steps without adversarial training. Every 5 steps, we update the weights of the discriminator and add a weight of 0.2 to LinDiff's adversarial loss to speed up the training process.

For the subjective evaluation of our system, we employ mean opinion scores (MOS) to assess the naturalness of the generated speech. The MOS were rated on a 1-to-5 scale and we report them along with the 95\% confidence intervals (CI). In addition, we perform objective evaluations using several metrics, including Mel-cepstrum distortion~\citep{kubichek1993mel} (MCD), error rate of voicing/unvoicing flags (V/UV), and correlation factor of F0 (F0 CORR) between the synthesized speech and the ground truth. To explore the diversity between the generated and real speeches, we calculate the Number of Statistically-Different Bins (NDB) and JensenShannon divergence (JSD). Furthermore, we evaluate the inference speed of our system on a single NVIDIA GeForce RTX 3090 GPU using the real-time factor (RTF).

\subsection{Comparison with other models}
We compared the proposed model in audio quality, diversity and sampling speed with other speech synthesis model, including 1) WaveNet\citep{oord2016wavenet}, an autoregressive generative model. 2) WaveGlow\citep{prenger2019waveglow}, a flow-based model. 3) HIFI-GAN V1\citep{kong2020hifi}, a GAN-based model. 4) WaveGrad\citep{chen2020wavegrad} and FastDiff\citep{huang2022fastdiff}, recently proposed DDPMs-based model.

The audio quality and sampling speed results are presented in Table \ref{tb1}. Our proposed method demonstrates the ability to synthesize high-fidelity speech with a limited number of steps. Even with just 3 steps, our model can generate speech of comparable quality to that produced by the autoregressive model, WaveNet \citep{oord2016wavenet}. However, the inference speed surpasses that of WaveNet and other conventional vocoders. In fact, its speed is on par with HIFI-GAN while outperforming it in terms of the quality of generated samples. Regarding sample diversity, Table \ref{tb2} shows that although LinDiff still lags behind the autoregressive model, WaveNet, it achieves greater variety in the generated speeches compared to other conventional vocoders.

\begin{table}
  \caption{Comparison with other convential nerual vocoders in terms of quality and synthesis speed with the model trained on single-speaker dataset, LJSpeech.}
  \label{tb1}
  \centering
  \begin{tabular}{c|cccc|c}
    \toprule
    & &\multicolumn{2}{c}{\textbf{Quality}}&  &\textbf{Speed }                \\

    Model     & MOS ($\uparrow$)    & MCD ($\downarrow$)& V/UV ($\downarrow$) &F0 CORR ($\uparrow$) &RTF ($\downarrow$)\\
    \midrule
    GT & 4.47$\pm$0.07  & $/$ &$/$  &$/$  &$/$     \\
    \midrule
    WaveNet (MOL) & \textbf{4.23$\pm$0.06}  & \textbf{1.74}&\textbf{6.87\%}&\textbf{0.89}  & 91.27 \\
    WaveGlow     & 3.79$\pm$0.09 & 2.83&17.18\% &0.68   &0.049  \\
    HIFI-GAN V1     & 3.94$\pm$0.08       & 2.08 &8.98\% &0.77 & \textbf{0.003}\\
    \midrule
    WaveGrad (noise schedule) &3.88$\pm$0.07 & 2.76&10.31\% &0.69&0.051\\
    FastDiff (4 steps) &4.05$\pm$0.07 & 2.56&8.59\% &0.79&0.025\\
    \midrule
    LinDiff (1 steps) &3.99$\pm$0.06 & 2.17 &9.12\% &0.74 & 0.004\\
    LinDiff (3 steps) &4.12$\pm$0.07 & 1.96 &8.75\% &0.79 & 0.013\\
    LinDiff (100 steps) &4.18$\pm$0.05 & 1.92 &7.78\% &0.82 & 0.520\\
    \bottomrule
  \end{tabular}
\end{table}

\begin{table}
  \caption{Comparison with other convential nerual vocoders in terms of diversity with the model trained on single-speaker dataset, LJSpeech.}
  \label{tb2}
  \centering
  \begin{tabular}{c|cccc|c}
    \toprule
    Model     &NDB ($\downarrow$)    & JSD ($\downarrow$)\\
    \midrule
    GT & $/$  & $/$ \\
    \midrule
    WaveNet (MOL) & \textbf{42}  &\textbf{0.002} \\
    WaveGlow     & 115 & 0.011      \\
    HIFI-GAN V1     & 69       & 0.004  \\
    \midrule
    WaveGrad (noise schedule) &122 &0.007\\
    FastDiff (4 steps) &65 & 0.005\\
    \midrule
    LinDiff (1 steps) &81 & 0.005 \\
    LinDiff (3 steps) &71 & 0.004 \\
    LinDiff (100 steps) &58 & 0.004 \\
    \bottomrule
  \end{tabular}
\end{table}

\subsection{Zero-shot experiment}
To further investigate the capabilities of our model, we trained it on the multi-speaker dataset, LibriTTS. We evaluated it using the Mel-spectrograms extracted from LJSpeech. Since the speakers in LJSpeech were not used during the training of this experiment, this task is referred to as Zero-shot speech generation. The results are presented in table \ref{tb3} below. It is evident from the table that all models perform worse when tasked with generating speech from multiple speakers. Our proposed transformer-based model suffers a significant performance drop in this scenario. The reason may be, in the case of a single-speaker dataset, the token embedding space is smaller than that of a multi-speaker dataset, as we split the input audio into small patch tokens. Therefore, modeling the latter is more challenging.
\begin{table}[t]
  \caption{Comparison with other convential nerual vocoders in terms of speech quality with the model trained on multi-speaker dataset. Zero-Shot MOS means we test the model on speakers that is not included in the training set.}
  \label{tb3}
  \centering
  \begin{tabular}{c|cc}
    \toprule
    Model     &MOS ($\uparrow$,Seen speaker) &Zero-Shot  MOS ($\uparrow$)\\
    \midrule
    GT & 4.47$\pm$0.07   &  4.47$\pm$0.07 \\
    \midrule
    WaveNet (MOL) & \textbf{4.10$\pm$0.08}  & \textbf{4.03$\pm$0.07}\\
    WaveGlow     & 3.64$\pm$0.06    &  3.56$\pm$0.07\\
    HIFI-GAN V1     & 3.94$\pm$0.08    &  3.89$\pm$0.06    \\
    \midrule
    WaveGrad (noise schedule) &3.74$\pm$0.07 &  3.71$\pm$0.06 \\
    FastDiff (4 steps) &3.95$\pm$0.07 &3.90$\pm$0.08\\
    \midrule
    LinDiff (1 steps) &3.73$\pm$0.08 &3.68$\pm$0.07\\
    LinDiff (3 steps) &3.83$\pm$0.08 &3.71$\pm$0.08\\
    LinDiff (100 steps) &3.92$\pm$0.07 &3.85$\pm$0.06\\
    \bottomrule
  \end{tabular}
\end{table}

\subsection{Ablation study}
\begin{figure}[!h]
  \centering
  \includegraphics[width=14cm]{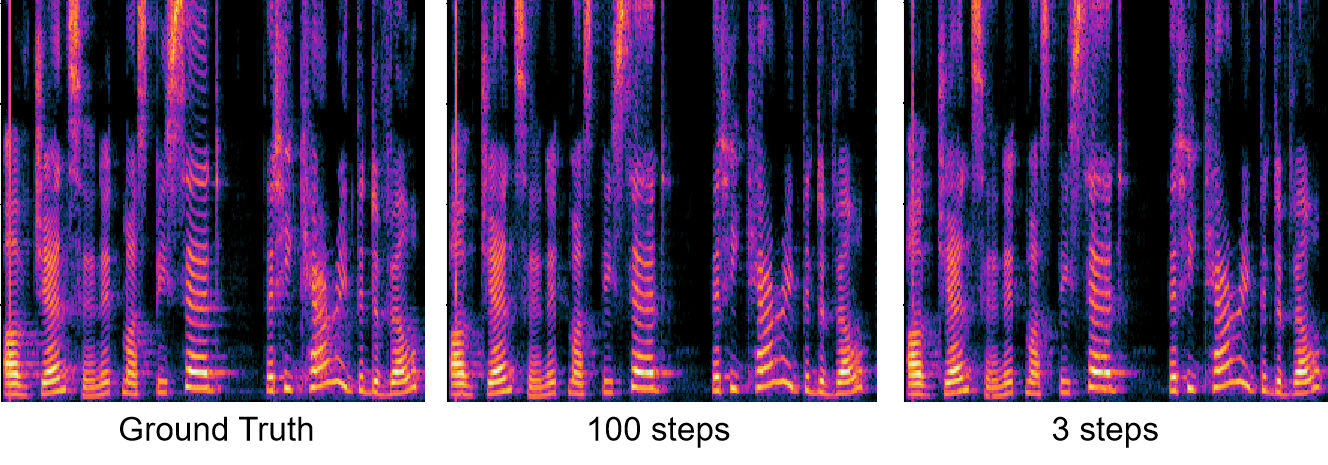}
  \caption{Visualization of spectrograms from the ground truth audio and predicted audio with 100 steps and 3 steps.}
  \label{fig2}
\end{figure}

\begin{figure}[!h]
  \centering
  \includegraphics[width=9cm]{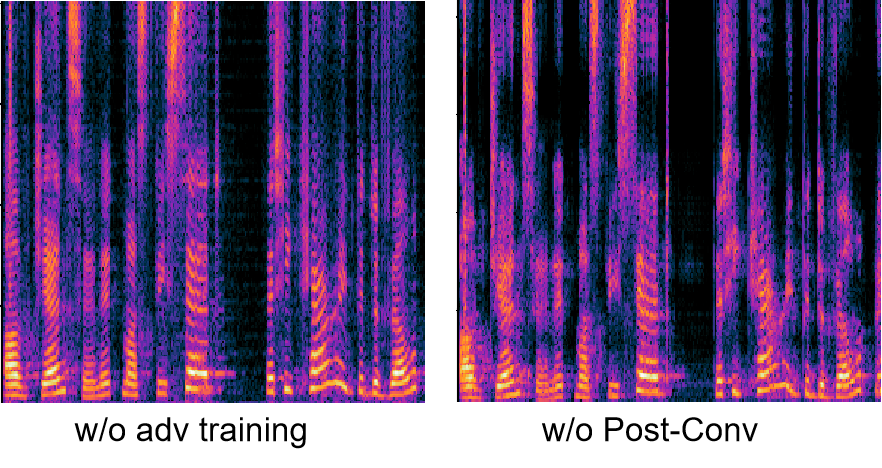}
  \caption{Visualization of spectrograms from predicted audio from the model removing specified module. w/o here means without. Diffusion steps were 3 in this experiment to demonstrate that the proposed Linear diffusion algorithm can synthesis relatively high-quality samples with limited steps even without adversarial training.}
  \label{fig3}
\end{figure}

We conducted separate experiments to evaluate the performance of our model after removing the Post conv layer or discarding adversarial training. The results demonstrate a significant performance decrease when either of these components is missing, with a notable increase in the presence of unwanted noise and artifacts in the generated audio. This is particularly evident in the spectrogram analysis of synthesized speech, which exhibits harmonic noise patterns as distinct horizontal lines when either component is removed. It can be found in Fig \ref{fig3}. In addition, we explore different sampling steps' influence on the final results. We compare the synthesized audio with 1 step and 100steps. Fig \ref{fig2} shows the results.
\begin{table}[!h]
  \caption{Ablation study results. Comparison of models with different configs. We set sampling steps to 100 in this experiment.}
  \label{tb4}
  \centering
  \begin{tabular}{c|cccc}
    \toprule
    Model     &MOS ($\uparrow$) &MCD ($\downarrow$)&V/UV ($\downarrow$)&F0 CORR ($\uparrow$)\\
    \midrule
    LinDiff origin (patch 64) & \textbf{4.18$\pm$0.05} &\textbf{1.92}&\textbf{7.78\%}&\textbf{0.82}\\
    \midrule
    LinDiff w/o Post-Conv     & 3.84$\pm$0.06 & 2.98&9.36\%&0.72  \\
    LinDiff w/o adv training  & 3.74$\pm$0.08 &2.91&10.39\%&0.76\\
    \midrule
    LinDiff (patch 128)     & 3.65$\pm$0.08 &2.87&16.57\%&0.67\\
    LinDiff (patch 256)     & 3.34$\pm$0.05  &3.34&17.89\%&0.59\\
    \bottomrule
  \end{tabular}
\end{table}
We also investigate the impact of patch size on our model. The results, as presented in Table \ref{tb4}, confirm that smaller patch sizes indeed yield higher quality samples. However, considering sampling efficiency, it is advisable to choose a sufficiently large patch size. After careful consideration, we select a patch size of 64 as it strikes the best balance between sample quality and sampling speed.

\section{Limitations}
As LinDiff leverages the powerful long-distance in-context modeling capability of the Transformer architecture, it also exhibits certain limitations.  Firstly, the computational cost of the Transformer increases quadratically with sequence length, making it challenging to generate very long speeches. In this experiment, the maximum length of the Transformer was set to 3600, which corresponds to approximately 10 seconds of synthesized speech. Secondly, the audio patch token in the LinDiff model exhibits more diversity in multi-speaker datasets, leading to inferior performance compared to models that solely rely on Convolutional Neural Networks (CNNs) as their backbone. 

\section{Conclusion}
In this work, we present LinDiff, a novel conditional diffusion model designed for fast and high-fidelity speech synthesis. By leveraging an Ordinary Differential Equation (ODE), we construct a diffusion path that offers improved fitting capabilities with reduced sampling steps compared to previous DDPMs. Moreover, LinDiff incorporates Transformer and CNN architectures. The Transformer model captures global information at a coarse-grained level, while the Convolutional layers handle fine-grained details. Additionally, we employ generative adversarial training to further enhance sampling speed and improve the quality of synthesized speech. Experimental results demonstrate that the proposed method can synthesis speech of comparable quality to autoregressive models, with a Real-Time Factor (RTF) of 0.013, making it significantly faster than real-time usage.

\newpage
{\small
\bibliographystyle{ieee_fullname}
\bibliography{egbib}
}

\end{document}